\documentclass[12pt]{article}
\usepackage{amsmath}
\usepackage{amssymb}
\usepackage{amstext}
\usepackage{amscd}
\usepackage{amsfonts}
\DeclareMathAlphabet{\EuFrak}{U}{euf}{m}{n}
\DeclareMathAlphabet{\EuScript}{U}{eus}{m}{n}
\newcommand{\be}{\begin{equation}}
\newcommand{\nd}{\noindent}
\newcommand{\ee}{\end{equation}}
\newcommand{\ben}{\begin{eqnarray}}
\newcommand{\een}{\end{eqnarray}}

\title{{\bf A direct proof of Jauregui-Tsallis' conjecture}}

\author{ A. Plastino and M. C. Rocca \\
Departamento de F\'{\i}sica-CCT-IFLP- CONICET \\
Fac. de Ciencias Exactas,
Universidad Nacional de La Plata\\
C.C. 67 (1900) La Plata, Argentina}

\date{December 1, 2010}

\begin{document}

\maketitle

\begin{abstract}

We give here  direct proof of a recent conjecture of Jauregui and
Tsallis about a new representation of Dirac's delta distribution
by means of q-exponentials. The proof is based in the use of
tempered ultradistributions' theory.

\end{abstract}

\newpage

\renewcommand{\theequation}{\arabic{section}.\arabic{equation}}

\section{Introduction}

Empirical analysis suggests that power-law behavior in the
(observed) distribution of some quantity is quite frequent in
nature. Indeed,  systems statistically described by {\it power-law
probability distributions} (PLD)
 are rather ubiquitous \cite{cero} and thus
of perennial interest \cite{uno,vign1,vign2}. Critical phenomena
are just a conspicuous example \cite{goldenfeld}.  Many objects
that come in different sizes have a self-similar power-law
distribution of their relative abundance over large size-ranges,
from cities to words to meteorites \cite{cero}. More specifically,
one often confronts a particularly important scenario: {\it
measuring real data distributed according to a q-Gaussian
probability law}, a special kind of power-law probability
distribution function (PDF), that is, a power-law distribution
that maximizes the so-called Tsallis' information measure $H_q$
under variance constraint, with

\be H_{q}\left( f\right) =\frac{1}{1-q}\left(
1-\int_{-\infty}^{+\infty} f(x)^{q} dx\right). \label{dino} \ee
Intense activity revolves around this information measure, with
hundreds of papers devoted to the study of its properties and
development of  applications in diverse scientific fields (see for
instance, as a small sample,
\cite{vign1,vign2,dino,super,gellmann,lissia,fromgibbs}, and
references therein) .
 The present effort is to be included in such context.

Tsallis and Jauregui have recently conjectured that, via
 probability distributions that maximize $H_q$ (called q-exponential
 functions) an interesting
representation of the Dirac Delta distribution $\delta_q(x)$ can
be given \cite{tsallis}. However, they could not prove their
conjecture and used numerical experiments that suggest its
validity.  In the wake of this achievement, Chevreuil, Plastino and Vignat
\cite{pladelta} provided a rigorous mathematical approach to the
problem and proved the conjecture by recourse to the notion of
superstatistics. Here we
 tackle again the subject and present a direct, structurally simpler proof by
 appealing to tempered ultradistributions' theory.

\section{The q-exponential function}

Statistical Mechanics' most notorious and renowned probability
distribution is that deduced by Gibbs for the canonical ensemble
\cite{reif, pathria}, usually referred to as the Boltzmann-Gibbs
equilibrium distribution

\be \label{gibbs}  p_G(i) =\frac{\exp{(-\beta E_i)}}{Z_{BG}}, \ee
with $E_i$ the  energy of the microstate labeled by $i$,
$\beta=1/k_B T$ the inverse temperature, $k_B$ Boltzmann's
constant, and $Z_{BG}$ the partition function. The exponential
term $F_{BG}=\exp{(-\beta E)}$ is called the Boltzmann-Gibbs
factor. Recently Beck and Cohen \cite{super} have advanced a
generalization, called ``superstatistics",  of this BG factor,
assuming that the inverse temperature $\beta$ is a stochastic
variable. The generalized statistical factor $F_{GS}$ is thus
obtained as the multiplicative convolution \be \label{bc}
F_{GS}=\int_0^\infty\, \frac{d\beta}{\beta}\,f(\beta)\,
\exp{(-\beta E)},\ee where $f(\beta)$ is the density probability
of the inverse temperature.  As stated above, $\beta$ is the
inverse temperature, but the integration variable may also be any
convenient intensive parameter. Superstatistics, meaning
``superposition of statistics", takes into account fluctuations of
such intensive parameters.

\nd Beck and Cohen also show that if $f(\beta)$ is a Gamma
distribution,  a special kind of thermostatistics arises, called
nonextensive thermostatistics,  a very active field, with
applications to several scientific disciplines
\cite{gellmann,lissia,fromgibbs,tsallis}. In working in a
nonextensive framework,  one has to deal with power-law
distributions called q-Gaussians , that maximize Tsallis'
information measure subject to appropriate constraints, with
$q\ne1$  a real positive parameter called the nonextensivity
index. More precisely, in the case of the celebrated canonical
distribution, there is only one constraint, the energy $E$, i.e.,
$\langle E\rangle = K\,\,\left(K\,\,\text{a positive
constant}\right)$ and  the equilibrium
 canonical distribution writes

\[
f_{q}(x)=\frac{1}{Z_{q}}\left(1-(1-q)\beta_q
E\right)_{+}^{\frac{1}{1-q}},
\]
with $(x)_{+}=\max\left(0,x\right)$ and $\beta_q$ and $Z_q$
standing for the nonextensive counterparts of $\beta$ and $Z_{BG}$
above. Defining the $q-$exponential function as

\begin{equation}
\label{qexp} e_{q}\left(x\right)=\left(1+ \left(1-q\right) x
\right)_{+}^{\frac{1}{1-q}}
\end{equation}
allows to rewrite the equilibrium distribution in the more natural
way
\[
f_{q}\left(x\right)=\frac{1}{Z_{q}} e_{q}\left(-\beta E \right).
\]
It is a classical result that as $q\rightarrow1,$   Tsallis
entropy reduces to Shannon's entropy \be \label{shannon}
H_{1}\left( f\right) =-\int_{-\infty}^{+\infty}f(x)\log f(x). \ee
Accordingly, the $q-$exponential function converges to the usual
exponential function.

\setcounter{equation}{0}

\section{Proof of Jauregui-Tsallis' conjecture}

\subsection{Preliminaries}

The problems of characterizing analytic functions whose boundary
values are elements of the spaces of distributions, or,
conversely, of finding representations of elements
of the quoted spaces of generalized functions by analytic
functions have a long history. Numerous papers have been written
concerning ultradistribution spaces of Sebastiao e Silva \cite{tp1}.
Such spaces are related to
the solvability and the regularity problems of partial
differential equations. Because of this relation, the study of the
structural problems as well as problems of various operations and
integral transformations in this setting is interesting in itself.

Thus, an analysis of spaces of distributions considered as
boundary values of analytic functions having appropriate growth
estimates, is of great value. One wishes to deal, in particular,
with the Dirac's integral representation in ultradistribution
spaces, with the convolution of tempered ultradistributions and
ultradistributions of exponential type (in Quantum Field Theory),
and with the integral transforms of tempered ultradistributions,
of which the best known is the Fourier complex transformation.
Recourse to such stuff will pave the way for our proof below. Some
explanatory material that should help to understand this proof is
given in Appendix I and Appendix II.

\subsection{Proof}
Our starting point is to consider $e_q(ikx)$ for $1<q<2$ and  $k$
a real number:
\begin{equation}
\label{ep4.1} e_q(ikx)=[1+i(1-q)kx]^{\frac {1} {1-q}}
\end{equation}

Then $e_q(ikx)$ is the cut along the real k-axis of the tempered
ultradistribution:
\begin{equation}
\label{ep4.2}
E_q(ikx)=\left\{H(x)H[\Im(k)]-H(-x)H[-\Im(k)]\right\}
[1+i(1-q)kx]^{\frac {1} {1-q}}
\end{equation}
where $H(x)$ is the Heaviside's step function. In (\ref{ep4.2})
$k$ is a complex variable.

According to \cite{tt3} we have the following formula:
\begin{equation}
\label{ep4.3} \int\limits_0^{\infty} \frac {x^{\mu-1}} {(1+\beta
x)^{\nu}}\;dx= {\beta}^{-\mu}\frac {\Gamma(\mu)\Gamma(\nu-\mu)}
{\Gamma(\nu)}
\end{equation}
$0<\Re(\mu)<\Re(\nu), |arg\beta|<\pi$, from which we deduce:
\begin{equation}
\label{ep4.4} \int\limits_0^{\infty} (x+\beta)^{\nu}\;dx= -\frac
{{\beta}^{1+\nu}} {1+\nu}
\end{equation}
$\Re(\nu)<-1, |arg\beta|<\pi$ Let $F_q(k)$ be given by:
\begin{equation}
\label{ep4.5} F_q(k)=\int\limits_{-\infty}^{\infty}E_q(ikx)\;dx
\end{equation}
Then:
\[F_q(k)=H[\Im(k)][(1-q)ik]^{\frac {1} {1-q}}
\int\limits_0^{\infty}\left[x+\frac {1} {(1-q)ik}\right]^{\frac
{1} {1-q}}\;dx-\]
\begin{equation}
\label{ep4.6} H[-\Im(k)][(q-1)ik]^{\frac {1} {1-q}}
\int\limits_0^{\infty}\left[x+\frac {1} {(q-1)ik}\right]^{\frac
{1} {1-q}}\;dx
\end{equation}
Using (\ref{ep4.4}) we obtain for (\ref{ep4.6}) ($1<q<2$)
\[F_q(k)=-H[\Im(k)]\frac {1} {(2-q)ik}-H[-\Im(k)]\frac {1} {(2-q)ik}=\]
\begin{equation}
\label{ep4.7} \frac {1} {2-q} \left(-\frac {1} {ik}\right)=\frac
{2\pi} {2-q} \delta(k)
\end{equation}
We reach an important milestone here. {\it Formula (\ref{ep4.7})
is the proof of Jauregui-Tsallis' conjecture for $k$ complex.}.

\vskip 3mm
 \nd The idea is now to consider
 the real axis in the variable $k$. To this effect  we note that
\[\frac {2\pi} {2-q}\phi(0)=\oint\limits_{\Gamma}F_q(k)\phi(k)\;dk=\]
\[\int\limits_{-\infty}^{\infty}\lim_{\epsilon\rightarrow 0^+}
\left\{\int\limits_0^{\infty}[1+(1-q)i(k+i\epsilon)x]^{\frac {1}
{1-q}}\;dx +\right.\]
\begin{equation}
\label{ep4.8}
\left.\int\limits_{-\infty}^0[1+(1-q)i(k-i\epsilon)x]^{\frac {1}
{1-q}}\;dx\right\} \phi(k)\;dk
\end{equation}
where $\phi(k)$ is an analytic  test function rapidly decreasing
(See Appendix II) . From (\ref{ep4.8}) we obtain:
\[\int\limits_{-\infty}^{\infty}e_q(ikx)\;dx=
\lim_{\epsilon\rightarrow 0^+}
\left\{\int\limits_0^{\infty}[1+(1-q)i(k+i\epsilon)x]^{\frac {1}
{1-q}}\;dx +\right.\]
\begin{equation}
\label{ep4.9}
\left.\int\limits_{-\infty}^0[1+(1-q)i(k-i\epsilon)x]^{\frac {1}
{1-q}}\;dx\right\}= \frac {2\pi} {2-q} \delta(k)
\end{equation}
We have reached our goal. Formula (\ref{ep4.9}) is the proof of
Jauregui-Tsallis' conjecture on the real axis.

\section{Conclusions}
In this paper we have straightforwardly proved the
Jauregui-Tsallis' conjecture by recourse to  tempered
ultradistributions' theory, on the complex plane and on the real
axis. We remark on the fact that the vital ingredient here are the
so-called q-exponential distributions, that play a significant
role in statistical mechanics as maximizers of Tsallis' entropy
under variance constraint. Thus we are in a rather curious
position of having obtained the proof of a mathematical conjecture
inspired by thermodynamic's ideas.

\newpage

\section{Appendix I: Distributions of Exponential Type}

\setcounter{equation}{0}

For the reader's benefit  we  briefly review the main properties
of Tempered Ultradistributions.

\nd {\bf Notations}. The notations are almost textually taken from
Ref. \cite{tp2}. Let $\boldsymbol{{\mathbb{R}}^n}$ (res.
$\boldsymbol{{\mathbb{C}}^n}$) be the real (resp. complex)
n-dimensional space whose points are denoted by
$x=(x_1,x_2,...,x_n)$ (resp $z=(z_1,z_2,...,z_n)$). We shall use
the notations:

(i) $x+y=(x_1+y_1,x_2+y_2,...,x_n+y_n)$\; ; \;
    $\alpha x=(\alpha x_1,\alpha x_2,...,\alpha x_n)$

(ii)$x\geqq 0$ means $x_1\geqq 0, x_2\geqq 0,...,x_n\geqq 0$

(iii)$x\cdot y=\sum\limits_{j=1}^n x_j y_j$

(iV)$\mid x\mid =\sum\limits_{j=1}^n \mid x_j\mid$

\nd Let $\boldsymbol{{\mathbb{N}}^n}$ be the set of n-tuples of
natural numbers. If $p\in\boldsymbol{{\mathbb{N}}^n}$, then
$p=(p_1, p_2,...,p_n)$, and $p_j$ is a natural number, $1\leqq
j\leqq n$. $p+q$ denote $(p_1+q_1, p_2+q_2,..., p_n+q_n)$ and
$p\geqq q$ means $p_1\geqq q_1, p_2\geqq q_2,...,p_n\geqq q_n$.
$x^p$ means $x_1^{p_1}x_2^{p_2}... x_n^{p_n}$. We shall denote by
$\mid p\mid=\sum\limits_{j=1}^n  p_j $ and by $D^p$ we denote the
differential operator
${\partial}^{p_1+p_2+...+p_n}/\partial{x_1}^{p_1}
\partial{x_2}^{p_2}...\partial{x_n}^{p_n}$

\nd For any natural $k$ we define $x^k=x_1^k x_2^k...x_n^k$ and
${\partial}^k/\partial x^k= {\partial}^{nk}/\partial x_1^k\partial
x_2^k...\partial x_n^k$

\nd The space $\boldsymbol{{\cal H}}$  of test functions such that
$e^{p|x|}|D^q\phi(x)|$ is bounded for any p and q is defined (
ref.\cite{tp2} ) by means of the countably set of norms:
\begin{equation}
\label{ep2.1}
{\|\hat{\phi}\|}_p=\sup_{0\leq q\leq p,\,x}
e^{p|x|} \left|D^q \hat{\phi} (x)\right|\;\;\;,\;\;\;p=0,1,2,...
\end{equation}
According to reference\cite{tp5} $\boldsymbol{{\cal H}}$  is a
$\boldsymbol{{\cal K}\{M_p\}}$ space
with:
\begin{equation}
\label{ep2.2}
M_p(x)=e^{(p-1)|x|}\;\;\;,\;\;\; p=1,2,...
\end{equation}
$\boldsymbol{{\cal K}\{e^{(p-1)|x|}\}}$ satisfies condition
$\boldsymbol({\cal N})$
of Guelfand ( ref.\cite{tp4} ). It is a countable Hilbert and nuclear
space:
\begin{equation}
\label{ep2.3}
\boldsymbol{{\cal K}\{e^{(p-1)|x|}\}} =\boldsymbol{{\cal H}} =
\bigcap\limits_{p=1}^{\infty}\boldsymbol{{\cal H}_p}
\end{equation}
where $\boldsymbol{{\cal H}_p}$ is obtained by completing
$\boldsymbol{{\cal H}}$ with the norm induced by
the scalar product:
\begin{equation}
\label{ep2.4}
{<\hat{\phi}, \hat{\psi}>}_p = \int\limits_{-\infty}^{\infty}
e^{2(p-1)|x|} \sum\limits_{q=0}^p D^q \overline{\hat{\phi}} (x) D^q
\hat{\psi} (x)\;dx \;\;\;;\;\;\;p=1,2,...
\end{equation}
where $dx=dx_1\;dx_2...dx_n$

\nd If we take the usual scalar product:
\begin{equation}
\label{ep2.5}
<\hat{\phi}, \hat{\psi}> = \int\limits_{-\infty}^{\infty}
\overline{\hat{\phi}}(x) \hat{\psi}(x)\;dx
\end{equation}
then $\boldsymbol{{\cal H}}$, completed with (\ref{ep2.5}), is the Hilbert space
$\boldsymbol{H}$
of square integrable functions.

\nd The space of continuous linear functionals defined on
$\boldsymbol{{\cal H}}$ is the space
$\boldsymbol{{\Lambda}_{\infty}}$ of the distributions of the
exponential type ( ref.\cite{tp2} ).

\nd The ``nested space''
\begin{equation}
\label{ep2.6}
{\Large{H}}=
\boldsymbol{(}\boldsymbol{{\cal H}},\boldsymbol{H},
\boldsymbol{{\Lambda}_{\infty}} \boldsymbol{)}
\end{equation}
is a Guelfand's triplet ( or a Rigged Hilbert space \cite{tp4} ).

\nd In addition we have: $\boldsymbol{{\cal
H}}\subset\boldsymbol{{\cal S}}
\subset\boldsymbol{H}\subset\boldsymbol{{\cal S}^{'}}\subset
\boldsymbol{{\Lambda}_{\infty}}$, where $\boldsymbol{{\cal S}}$ is
the Schwartz space of rapidly decreasing test functions
(ref\cite{tp6}).

\nd Any Guelfand's triplet
${\Large{G}}=\boldsymbol{(}\boldsymbol{\Phi},
\boldsymbol{H},\boldsymbol{{\Phi}^{'}}\boldsymbol{)}$ has the
fundamental property that a linear and symmetric operator on
$\boldsymbol{\Phi}$, admitting an extension to a self-adjoint
operator in $\boldsymbol{H}$, has a complete set of generalized
eigen-functions in $\boldsymbol{{\Phi}^{'}}$ with real
eigenvalues.

\section{Appendix II: Tempered Ultradistributions}
\setcounter{equation}{0}

The Fourier transform of a function $\hat{\phi}\in \boldsymbol{{\cal H}}$
is
\begin{equation}
\label{ep3.1}
\phi(z)=\frac {1} {2\pi}
\int\limits_{-\infty}^{\infty}\overline{\hat{\phi}}(x)\;e^{iz\cdot x}\;dx
\end{equation}
$\phi(z)$ is entire analytic and rapidly decreasing on straight lines
parallel
to the real axis. We shall call $\boldsymbol{{\EuFrak H}}$
the set of all such functions.
\begin{equation}
\label{ep3.2}
\boldsymbol{{\EuFrak H}}={\cal F}\left\{\boldsymbol{{\cal H}}\right\}
\end{equation}
It is a $\boldsymbol{{\cal Z}\{M_p\}}$ space ( ref.\cite{tp5} ),
countably normed and complete, with:
\begin{equation}
\label{ep3.3}
M_p(z)= (1+|z|)^p
\end{equation}
$\boldsymbol{{\EuFrak H}}$ is also a nuclear space with norms:
\begin{equation}
\label{ep3.4}
{\|\phi\|}_{pn} = \sup_{z\in V_n} {\left(1+|z|\right)}^p
|\phi (z)|
\end{equation}
where $V_k=\{z=(z_1,z_2,...,z_n)\in\boldsymbol{{\mathbb{C}}^n}:
\mid Im z_j\mid\leqq k, 1\leqq j \leqq n\}$

\nd We can define the usual scalar product:
\begin{equation}
\label{ep3.5}
<\phi (z), \psi (z)>=\int\limits_{-\infty}^{\infty}
\phi(z) {\psi}_1(z)\;dz =
\int\limits_{-\infty}^{\infty} \overline{\hat{\phi}}(x)
\hat{\psi}(x)\;dx
\end{equation}
where:
\[{\psi}_1(z)=\int\limits_{-\infty}^{\infty}
\hat{\psi}(x)\; e^{-iz\cdot x}\;dx\]
and $dz=dz_1\;dz_2...dz_n$

\nd By completing $\boldsymbol{{\EuFrak H}}$ with the norm induced
by (\ref{ep3.5}) we get the Hilbert space of square integrable
functions.

\nd The dual of $\boldsymbol{{\EuFrak H}}$ is the space
$\boldsymbol{{\cal U}}$ of tempered ultradistributions (
ref.\cite{tp2} ). In other words, a tempered ultradistribution is
a continuous linear functional defined on the space
$\boldsymbol{{\EuFrak H}}$ of entire functions rapidly decreasing
on straight lines parallel to the real axis.

\nd The set ${\Large{U}}= \boldsymbol{({\EuFrak
H},H,{\cal U})}$ is also a Guelfand's triplet.

\nd Moreover, we have: $\boldsymbol{{\EuFrak
H}}\subset\boldsymbol{{\cal S}}
\subset\boldsymbol{H}\subset\boldsymbol{{\cal S}^{'}}\subset
\boldsymbol{{\cal U}}$.

$\boldsymbol{{\cal U}}$ can also be characterized in the following way
( ref.\cite{tp2} ): let $\boldsymbol{{\cal A}_{\omega}}$ be the space of
all functions $F(z)$ such that:

${\Large {\boldsymbol{I}}}$-
$F(z)$ is analytic for $\{z\in \boldsymbol{{\mathbb{C}}^n} :
|Im(z_1)|>p, |Im(z_2)|>p,...,|Im(z_n)|>p\}$.

${\Large {\boldsymbol{II}}}$-
$F(z)/z^p$ is bounded continuous  in
$\{z\in \boldsymbol{{\mathbb{C}}^n} :|Im(z_1)|\geqq p,|Im(z_2)|\geqq p,
...,|Im(z_n)|\geqq p\}$,
where $p=0,1,2,...$ depends on $F(z)$.

\nd Let $\boldsymbol{\Pi}$ be the set of all $z$-dependent
pseudo-polynomials, $z\in \boldsymbol{{\mathbb{C}}^n}$. Then
$\boldsymbol{{\cal U}}$ is the quotient space:

${\Large {\boldsymbol{III}}}$-
$\boldsymbol{{\cal U}}=\boldsymbol{{\cal A}_{\omega}/\Pi}$

\nd By a pseudo-polynomial we understand a function of $z$ of the
form $\;\;$ $\sum_s z_j^s G(z_1,...,z_{j-1},z_{j+1},...,z_n)$ with
$G(z_1,...,z_{j-1},z_{j+1},...,z_n)\in\boldsymbol{{\cal
A}_{\omega}}$

\nd Due to these properties it is possible to represent any
ultradistribution as ( ref.\cite{tp2} ):
\begin{equation}
\label{ep3.6}
F(\phi)=<F(z), \phi(z)>=\oint\limits_{\Gamma} F(z) \phi(z)\;dz
\end{equation}
$\Gamma={\Gamma}_1\cup{\Gamma}_2\cup ...{\Gamma}_n$
where the path ${\Gamma}_j$ runs parallel to the real axis from
$-\infty$ to $\infty$ for $Im(z_j)>\zeta$, $\zeta>p$ and back from
$\infty$ to $-\infty$ for $Im(z_j)<-\zeta$, $-\zeta<-p$.
( $\Gamma$ surrounds all the singularities of $F(z)$ ).

\nd Formula (\ref{ep3.6}) will be our fundamental representation
for a tempered ultradistribution. Sometimes use will be made of
``Dirac formula'' for ultradistributions (Ref. \cite{tp1}):
\begin{equation}
\label{ep3.7}
F(z)=\frac {1} {(2\pi i)^n}\int\limits_{-\infty}^{\infty}
\frac {f(t)} {(t_1-z_1)(t_2-z_2)...(t_n-z_n)}\;dt
\end{equation}
where the ``density'' $f(t)$ is such that
\begin{equation}
\label{ep3.8}
\oint\limits_{\Gamma} F(z) \phi(z)\;dz =
\int\limits_{-\infty}^{\infty} f(t) \phi(t)\;dt
\end{equation}
While $F(z)$ is analytic on $\Gamma$, the density $f(t)$ is in
general singular, so that the r.h.s. of (\ref{ep3.8}) should be interpreted
in the sense of distribution theory.

\nd Another important property of the analytic representation is
the fact that on $\Gamma$, $F(z)$ is bounded by a power of $z$
(Ref. \cite{tp2}):
\begin{equation}
\label{ep3.9}
|F(z)|\leq C|z|^p
\end{equation}
where $C$ and $p$ depend on $F$.

\nd The representation (\ref{ep3.6}) implies that the addition of
a pseudo-polynomial $P(z)$ to $F(z)$ do not alter the
ultradistribution:
\[\oint\limits_{\Gamma}\{F(z)+P(z)\}\phi(z)\;dz=
\oint\limits_{\Gamma} F(z)\phi(z)\;dz+\oint\limits_{\Gamma}
P(z)\phi(z)\;dz\]
But:
\[\oint\limits_{\Gamma} P(z)\phi(z)\;dz=0\]
as $P(z)\phi(z)$ is entire analytic in some of the variables $z_j$
(and rapidly decreasing),
\begin{equation}
\label{ep3.10} \therefore \;\;\;\;\oint\limits_{\Gamma}
\{F(z)+P(z)\}\phi(z)\;dz= \oint\limits_{\Gamma} F(z)\phi(z)\;dz.
\end{equation}

\end{document}